\documentclass[lineno]{jfm}

\usepackage{caption} 
\usepackage{pgfplots}
\usepackage{lineno}
\usepackage{setspace}
\usepackage{afterpage}
\usepackage{rotating}
\usepackage{mwe}
\usepackage{xcolor}
\usepackage{verbatim}
\usepackage{array}
\usepackage{graphicx}
\usepackage{newtxtext}
\usepackage{newtxmath}
\usepackage{natbib}
\usepackage{hyperref}
\hypersetup{
    colorlinks = true,
    urlcolor   = blue,
    citecolor  = black,
}

\newcommand{\RomanNumeralCaps}[1]

\usepackage{amssymb}
\usepackage{epstopdf, epsfig}
\usepackage{amsmath}%%% added
\usepackage{psfrag}  %%% added
\usepackage{subfigure} %%%added
\usepackage{mathrsfs}   %%%added
\usepackage{color}
\usepackage{overpic}
\usepackage{soul}
\usepackage{relsize}
\usepackage{xcolor, soul}  
\usepackage{pifont}
\usepackage{cleveref}
\newcommand{\averzt}[1]{\left\langle {#1} \right\rangle_{zt}}
\newcommand{\averyz}[1]{\left\langle {#1} \right\rangle}
\newcommand{\rme}{\mathrm{e}}
\newcommand{\abs}[1]{\left|#1\right|}
\newcommand{\ep}{\epsilon}
\newcommand{\R}{R_\lambda}
\newcommand{\iy}{\infty}

\newcommand{\kz}{\kappa_z}
\newcommand{\ky}{\kappa_y}
\newcommand{\kt}{\widetilde \kappa}

\newcommand{\xbar}{\overline{x}}
\newcommand{\ubar}{\overline{u}}
\newcommand{\vbar}{\overline{v}}
\newcommand{\wbar}{\overline{w}}

\newcommand{\ut}{\widetilde u}
\newcommand{\vt}{\widetilde v}
\newcommand{\wt}{\widetilde w}

\newcommand{\xt}{\widetilde x}

\newcommand{\ub}{\breve u}

\newcommand{\xb}{\breve x}

\shortauthor{Pierre Ricco}

\shorttitle{Scaling of boundary-layer disturbances}
\title{Scaling of boundary-layer disturbances exposed to free-stream turbulence}

\author{
    Pierre Ricco\corresp{\email{p.ricco@sheffield.ac.uk}} 
}

\affiliation{Department of Mechanical Engineering, The University of Sheffield, \\ Mappin Street, S1 3JD Sheffield, United Kingdom}
\begin{document}

\maketitle

\begin{abstract}
\textcolor{blue}{\textbf{Ricco, P. ``Scaling of boundary-layer disturbances exposed to free-stream turbulence'', {\em J. Fluid Mech.}, vol. 972, A3 (2023)}} \\
We present theoretical results related to the experimental findings of \cite{matsubara-alfredsson-2001} on the scaling of the energy spectra of the Klebanoff modes, i.e. streamwise-elongated vortical disturbances generated by free-stream turbulence in a flat-plate transitional boundary layer. The scaling is explained by a model that describes the streamwise evolution of the streamwise and spanwise energy spectra. The theoretical framework is based on the quasi-steady asymptotic solution of the boundary-region equations, on an axial-symmetric model of the free-stream spectrum, and on the spectral response of the boundary layer to the external perturbations.
\end{abstract}

\section{Introduction}

The transition of a boundary layer from the laminar regime to fully developed turbulence is a central problem in an immense range of technological applications because turbulent wall friction can be several times larger than that exerted by a laminar boundary layer. Frictional losses in the boundary layer are responsible for the performance degradation of engineering flow systems, such as turbomachinery and jet engines, for the enhanced aerodynamic drag of transport vehicles, and, in turn, for wasted fuel consumption, unwanted noise production and environmental pollution.
For design purposes, it is therefore paramount to be able to predict under which conditions boundary-layer transition occurs. Free-stream turbulence acts as a triggering factor for transition, and it has been shown that the transition Reynolds number decreases as the free-stream turbulence level increases \citep{mayle-1991}. 

\cite{dryden-1936} and \cite{taylor-1939} were probably the first to study the effects of free-stream turbulence on a flat-plate boundary layer. They showed that the dominant streamwise velocity fluctuations generated by free-stream turbulence in the boundary layer are of very low frequency and reach amplitudes that can be several times larger than those in the free stream.

The Dryden--Taylor observations did not receive much attention until \cite{klebanoff-1971} carried out experiments in which he reproduced the earlier findings of Dryden and Taylor. Klebanoff demonstrated that the disturbances grow more or less linearly with the boundary-layer thickness, and they are quite narrow in the spanwise direction. Klebanoff referred to these disturbances as `breathing modes' because, as noted earlier by \cite{taylor-1939}, they appeared to correspond to a thickening and thinning of the boundary layer. \cite{kendall-1991} renamed them Klebanoff modes, and that name has taken hold even though these disturbances are not modes in the strict mathematical sense, i.e. they are not homogeneous solutions of differential equations.

The early transition experiments were conducted at very low free-stream turbulence levels ($Tu$$<$0.1\%), but more recent experiments, such as those by \cite{westin-etal-1994}, \cite{matsubara-alfredsson-2001}, \cite{fransson-etal-2005}, \cite{fransson-shahinfar-2020} and \cite{mamidala-etal-2022} were carried out at higher turbulence levels. However, the results are invariably the same. The dominant streamwise velocity fluctuations are always of the Klebanoff type, i.e. the boundary layer acts as a low-frequency-pass filter on the free-stream perturbation spectrum, and amplifies streamwise stretched streaky vortical structures. The spanwise wavelength of the Klebanoff modes is constant along the streamwise direction, and the peak amplitude occurs at the same Blasius-similarity wall-normal location. Direct numerical simulations have also been employed to study the development of low-frequency streaks and the induced bypass transition \citep{jacobs-durbin-2001,ovchinnikov-choudhari-piomelli-2008,yao-etal-2022}.

The mathematical framework describing the incompressible Klebanoff modes was developed by \cite{leib-wundrow-goldstein-1999} (LWG99). They proved that these disturbances, near the leading edge, are well represented by forced solutions of the linearized unsteady boundary-layer equations for which the spanwise viscous effects are negligible. As the mean boundary layer grows downstream, these equations lose their validity because the spanwise length scale of the Klebanoff modes becomes comparable with the boundary-layer thickness. Their dynamics is then ruled by the unsteady boundary-region equations, i.e. the Navier-Stokes equations where the spanwise viscous ter.m.s. are retained, while the streamwise pressure gradient and the viscous effects can be neglected because the perturbations are of low frequency and streamwise elongated. The boundary-region equations, and their terminology, were first used by \cite{kemp-1951} to study the corner boundary-layer problem. A crucial ingredient in the LWG99 formulation is the continuous action of the free-stream perturbations that are responsible for the generation and evolution of the Klebanoff modes. LWG99 utilized matched asymptotic expansions to obtain the initial and outer boundary conditions that synthesize the interaction between the free-stream flow and the boundary-layer flow. \cite{wundrow-goldstein-2001} and \cite{ricco-etal-2011} extended the linearized study of LWG99 to include nonlinear effects, focusing on the steady and unsteady cases, respectively. \cite{ricco-etal-2011} also explained the occurrence of nonlinear effects in the results by \cite{matsubara-alfredsson-2001}, and studied the secondary instability of the saturated Klebanoff modes, thereby describing the mechanism at the heart of bypass transition induced by free-stream turbulence. Extensions to the compressible regime include the investigations by \cite{ricco-wu-2007}, \cite{ricco-tran-ye-2009}, \cite{ricco-shah-hicks-2013} and \cite{marensi-etal-2017}.

Other theories describing the Klebanoff modes have been proposed. The non-modal growth theory \citep{schmid-henningson-2001} and the optimal growth theory \citep{andersson-berggren-henningson-1999,luchini-2000} model the growth of streaky disturbances already present in the boundary layer, while allowing the disturbances to vanish in the free stream. Continuous Orr-Sommerfeld modes have also been used extensively since \cite{jacobs-durbin-2001} to synthesize the penetration of free-stream disturbances into a boundary layer. Reviews of this approach are found in \cite{dong-wu-2013}, \cite{ricco-etal-2016} and \cite{durbin-2017}.

In the present study, we develop the theoretical background of previously unexplained experimental results of a transitional boundary layer exposed to free-stream turbulence, reported by \cite{matsubara-alfredsson-2001} (MA01). These findings are remarkable because the energy spectra at different streamwise locations were found to collapse on one another when scaled properly. MA01 described their discovery as ``an unexpected new finding'' and their energy spectra showing ``an astonishing similarity'' for which ``there is no theoretical explanation''.

In \S\ref{sec:mat-alf-results}, the experimental findings of MA01 on the scaling of the Klebanoff modes are discussed. In \S\ref{sec:flow-system}, we present the key features of the mathematical framework describing the Klebanoff modes, while the theoretical results behind the experimental findings of MA01 are found in \S\ref{sec:results}. Section \ref{sec:conclusions} contains the conclusions.

%--------------------------------------------------------------------------
\section{Discussion of the experimental results of Matsubara \& Alfredsson}
\label{sec:mat-alf-results}

MA01 studied experimentally an incompressible flow of uniform velocity $U_\iy^*$ past a thin flat plate located in a low-speed wind tunnel. Rigid grids were placed upstream of the leading edge of the plate to generate free-stream vortical disturbances.  A thin laminar boundary layer developed over the flat plate and transitioned to a fully-developed turbulent boundary layer because of the perturbative action of the free-stream disturbances. The objective of the MA01 study was to fully characterize the transitional boundary layer.
In our discussion of the MA01 results and in the theoretical analysis, the flow is described through a Cartesian coordinate system, i.e. $\bf{x}^*$ = $x^* \hat{\bf{i}} + y^* \hat{\bf{j}} + z^* \hat{\bf{k}}$, where $x^*,y^*,z^*$ define the streamwise, wall-normal and spanwise directions, respectively, and the superscript $^*$ indicates a dimensional quantity. The flat plate is located at $y^*=0$, and its leading edge is at $x^*=0$. Lengths are scaled by $\Lambda_z^*$, the integral spanwise length scale of the free-stream vortical disturbances, velocities are scaled by $U_\iy^*$, pressure is scaled by $\rho^* U_\iy^{*2}$, where $\rho^*$ is the density, and time is scaled by $ \Lambda_z^*/U_\iy^*$. The kinematic viscosity is denoted by $\nu^*$. Non-dimensional quantities are not marked by any symbol. 

%-------------------------------------------
\subsection{Validity of linearized dynamics}

\begin{figure}
 \begin{center}
  \includegraphics[width=0.65\textwidth]{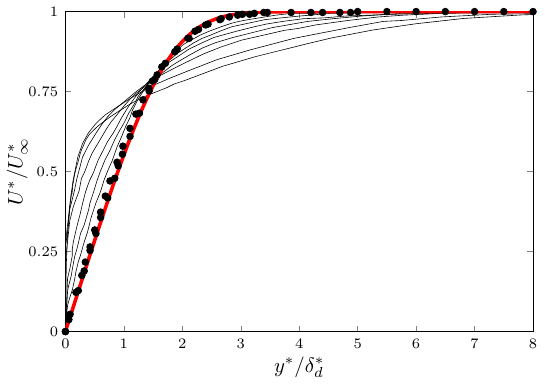}
  \caption{Mean boundary-layer streamwise velocity profiles reported by MA01 for $x^*$$\leq$$500$mm (black circles) and $700$mm$\leq$$x^*$$\leq$$1900$mm (thin lines). The red thick line denotes the numerical solution of the Blasius laminar boundary-layer flow and $\delta_d^*$ indicates the displacement thickness.}
  \label{fig:matsubara-figure-2}
 \end{center}
\end{figure}

As our theoretical framework hinges on the assumption that the boundary-layer disturbances are described by a linearized dynamics, we first examine the MA01 findings to support our hypothesis. Figure \ref{fig:matsubara-figure-2} shows the mean boundary-layer streamwise velocity profiles measured by MA01 at different streamwise locations. The data displayed by the black circles correspond to the three streamwise stations that are closest to the leading edge, i.e. $x^*$$=$100, 300, 500mm. The data represented by the thin lines were acquired at $x^*$$>$$500$mm. The black-circle data show excellent agreement with the numerical solution of the Blasius laminar boundary-layer flow, represented by the thick red line, while the thin-line data deviate progressively more and more from the laminar solution as $x^*$ increases. For $x^*$$>$$500$mm, nonlinear effects become important as the boundary-layer perturbations grow in amplitude, and the wall-shear stress is enhanced as fully-developed turbulence ensues. 
These results are evidence of the perturbed flow obeying a linearized dynamics at the locations closest to the leading edge because the mean-flow profiles follow the laminar solution. Figure 3 in MA01 further reveals that the boundary-layer thickness and the shape factor match the laminar values for $x^*$$\leq$$700$mm. Additional support for these results is given by profiles of the root-mean-square (r.m.s.) of the streamwise velocity fluctuations, shown in figure 2c of MA01, which denote clear signs of nonlinear effects for $x^*$$\geq$$1100$mm, such as the disturbances growing in the outer part of the boundary layer, and the perturbation peak moving closer to the wall. The theoretical and numerical results that match quantitatively the nonlinear MA01 data are discussed in \cite{ricco-etal-2011}. We conclude that a linearized dynamics can be utilized to study the perturbed flow for $x^*$$\leq$$500$mm, despite the free-stream turbulence intensity not being vanishingly small for these experiments, i.e. $Tu=2.2$\% (refer to grid A in table 1 in MA01).

%------------------------------------------------------
\subsection{Scaling of experimental turbulence spectra}
\label{sec:scaling-spectra}

\begin{figure}
 \begin{center}
  \begin{picture}(400,150)% 
   \includegraphics[width=\textwidth]{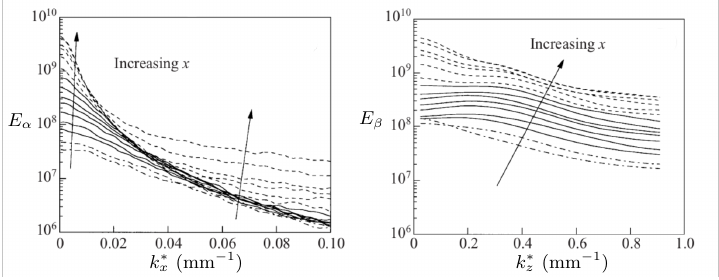}
   \put(-222,130){(a)}
   \put(-33,130){(b)}
  \end{picture}
  \begin{picture}(400,150)% 
   \includegraphics[width=\textwidth]{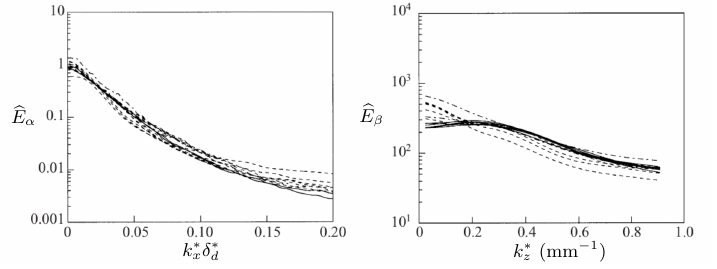}
   \put(-222,128){(c)}
   \put(-33,128){(d)}
  \end{picture}
  \caption{(a,b) Reproduction of figure 13 in MA01. Energy spectra as functions of (a) the streamwise wavenumber and (b) the spanwise wavenumber. (c,d) Reproduction of figure 14 in MA01. Rescaled energy spectra as functions of (c) the scaled streamwise wavenumber and (b) the dimensional spanwise wavenumber. Data were acquired at $y^*/\delta_d^*$$=$1.2. The solid lines are for $x^*$$=$120, 150, 200, 250, 300, 400, 500 mm. Labels in the original graphs have been changed to conform to the present notation.}
  \label{fig:matsubara-figure-13-14}
 \end{center}
\end{figure}
%-----------
Figures \ref{fig:matsubara-figure-13-14}(a,b), a reproduction of figure 13 in MA01, depict streamwise velocity energy spectra at $y^*/\delta_d^*=1.2$, where $\delta_d^*$ is the displacement thickness. For this experimental dataset, $U_\infty^*=5$m/s and $\Lambda_z^*=7$mm, computed from the autocorrelation of the streamwise velocity shown in figure 7 on p. 161 of MA01. The Reynolds number based on $\Lambda_z^*$ is $\R$$=$$U_\iy^* \Lambda^*_z/\nu^*$=2232. The spectrum $E_\alpha$ is shown as a function of the streamwise wavenumber $k_x^*=2 \pi f^*/U_\infty^*$, where $f^*$ is the frequency (figure \ref{fig:matsubara-figure-13-14}a), and the spectrum $E_\beta$ is shown as a function of the spanwise wavenumber $k_z^*=2 \pi /\lambda_z^*$, where $\lambda_z^*$ is the spanwise wavelength (figure \ref{fig:matsubara-figure-13-14}b).

The wavenumbers in figures \ref{fig:matsubara-figure-13-14}(a,b) are dimensional, while in our theoretical analysis they are scaled by $\Lambda_z^*$, that is, $k_x=k_x^* \Lambda_z^*$ and $k_z=k_z^* \Lambda_z^*$. The spectra $E_\alpha$ and $E_\beta$ are linked to the variance of the streamwise velocity fluctuations,

\begin{equation}
\label{eq:spectra}
\epsilon^2 \averzt{u^{\prime 2}}(x,y) = 
C_\alpha \int_0^\infty E_\alpha(k_x) \mathrm{d}k_x =
C_\beta \int_0^\infty E_\beta(k_z) \mathrm{d}k_z,
\end{equation}
where $C_\alpha$ and $C_\beta$ are constants, computed in \S\ref{sec:c-constants}, and $\averzt{\cdot}$ indicates averaging along $z$ and over $t$. In figure \ref{fig:matsubara-figure-13-14}, the dash-dotted lines refer to locations upstream of the solid lines, while the dashed lines correspond to locations downstream of the solid lines.

In figure \ref{fig:matsubara-figure-13-14}(a), for $x^*$$\leq$500 mm, the dash-dotted and solid lines show that the low-wavenumber portion of the spectrum grows downstream, while the high-wavenumber portion is unchanged. This behaviour confir.m.s. that the boundary layer acts as a low-frequency-pass filter \citep{durbin-2017}, consistently with the algebraic growth of the streamwise-elongated, low-frequency Klebanoff modes. The high-frequency free-stream disturbances do not penetrate sufficiently into the boundary layer to reach these wall-normal locations.
Nonlinear effects becomes predominant further downstream, where the high-wavenumber fluctuations grow more significantly than the low-wavenumber ones (dashed lines). Figure \ref{fig:matsubara-figure-13-14}(b) shows that the spanwise energy spectrum grows uniformly for all the spanwise wavenumbers.

Figures \ref{fig:matsubara-figure-13-14}(c,d) are a reproduction of figure 14 in MA01. The spectra $E_\alpha$ and $E_\beta$, shown in figures \ref{fig:matsubara-figure-13-14}(a,b), are scaled as (the symbol $\widehat{\cdot}$ is used here in lieu of $^*$ in MA01)
\begin{equation}
\label{eq:ealpha-ebeta-hats}
    \widehat E_\alpha = \frac{E_\alpha}{C_e Re_x^{3/2}}, \quad \widehat E_\beta = \frac{E_\beta}{C_e Re_x},
\end{equation}
where $Re_x=U_\infty^* x^*/\nu^*$, and the constant $C_e=16$ is the same for the two spectra. 
The scaling of $E_\beta$ with $Re_x$ is expected because the integral of $E_\beta$ along $k_z$, given by the last equation of \eqref{eq:spectra}, is equal to the variance of the streamwise velocity fluctuations, which grows linearly with $Re_x$, as shown by the experimental results in figure 2d of MA01. On the abscissas of figures \ref{fig:matsubara-figure-13-14}(c,d), the streamwise wavenumber is scaled by the displacement thickness $\delta_d^*$, while the spanwise wavenumber is dimensional. Both sets of profiles represented by the solid lines show excellent collapse when rescaled. The objective of our study is explain the scaling of those solid lines in figures \ref{fig:matsubara-figure-13-14}(c,d). 

This scaling demonstrates that the streamwise spectrum $E_\alpha$ grows downstream at a faster rate (proportional to $Re_x^{3/2}$) than its integral across the streamwise wavenumbers $\epsilon^2 \averzt{u^{\prime 2}}$,  which grows linearly with $Re_x$, as shown in figure 2d of MA01. The different growth rates are caused by the low-frequency fluctuations becoming larger more rapidly than the high-frequency ones, as shown in figure \ref{fig:matsubara-figure-13-14}(a). 

It is worth mentioning that \cite{zhigulev-uspenskii-ustinov-2009}, in their figures 7 and 8, reported similar scaling of streamwise spectra, in their case by $Re_x^2$ and $Re_x^{3/2}$, for different boundary-layer datasets collected in their low-turbulence wind tunnel ($Re_x^2$ and $Re_x^{3/2}$ were written as $\epsilon^2 \averzt{u^{*\prime 2}}$ and $\epsilon^2 \averzt{u^{*\prime 2}} \delta_d^*$, respectively, in their formulas (2.8) and (2.9)). They attributed the scaling by $Re_x^{3/2}$ to nonlinear effects. We show in the following that the scaling of the MA01 spectra can be explained by asymptotic results emerging from the linearized theory of LWG99, although our form of free-stream spectrum does model nonlinear effects through its streamwise dependency. 

%=======================================================
\subsection{Computation of ${C_\alpha}$ and ${C_\beta}$}
\label{sec:c-constants}

\begin{figure}
 \begin{center}
  \includegraphics[width=0.7\textwidth]{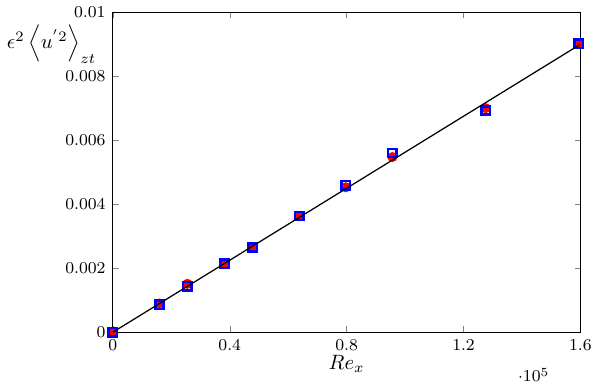}
  \caption{Growth of r.m.s. of streamwise velocity fluctuations as a function of $Re_x$, computed by integrating the spectra $E_\alpha$ (red circles) and  $E_\beta$ (blue squares) shown by solid lines in figures \ref{fig:matsubara-figure-13-14}(a,b).}
  \label{fig:c-alpha-c-beta}
 \end{center}
\end{figure}

The constants ${C_\alpha}$ and ${C_\beta}$ in \eqref{eq:spectra} are found as follows.
The integrals in \eqref{eq:spectra} are first computed by using the spectral data in figures \ref{fig:matsubara-figure-13-14}(a,b) at different streamwise locations $Re_x$. For the experimental data of figure \ref{fig:matsubara-figure-13-14}, MA01 do not report the values of $\epsilon^2 \averzt{u^{\prime 2}}$ at different $Re_x$. The data shown in figure 2d on p. 156 of MA01 for a similar set of flow conditions can, however, be used for our purpose because that graph shows that the r.m.s. of the streamwise velocity starts to deviate from the linear behaviour when it reaches a value of about $9 \cdot 10^{-3}$. 
The constants ${C_\alpha}$ and ${C_\beta}$ can thus be found by linear fitting of the integrated experimental data in order to obtain $\epsilon^2 \averzt{u^{\prime 2}}=9 \cdot 10^{-3}$ at $Re_x=159438$, which is the most downstream location where the data of figure \ref{fig:matsubara-figure-13-14} obey the scaling discussed in \S\ref{sec:scaling-spectra} (denoted by solid lines). Data downstream of this location, displayed by dashed lines in figures \ref{fig:matsubara-figure-13-14}(a,b), are affected by nonlinear effects, similarly to the r.m.s. data larger than $9 \cdot 10^{-3}$ in figure 2d on p. 156 of MA01.
The computed values are ${C_\alpha}=1.62 \cdot 10^{-10}$ and ${C_\beta}=4 \cdot 10^{-12}$. Figure \ref{fig:c-alpha-c-beta} shows that the r.m.s. values, obtained by integrating $E_\alpha$ and $E_\beta$, agree well with each other and grow linearly with $Re_x$ as expected. MA01 give the free-stream turbulence level for this experimental dataset, $Tu(\%)=0.022$, and we thus take $\epsilon=0.022$.

%=============================================================
\subsection{Power-law dependence of scaled turbulence spectra}
\label{sec:best-fitting}

\begin{figure}
 \begin{center}
 \begin{picture}(400,150)% 
   \includegraphics[width=0.49\textwidth]{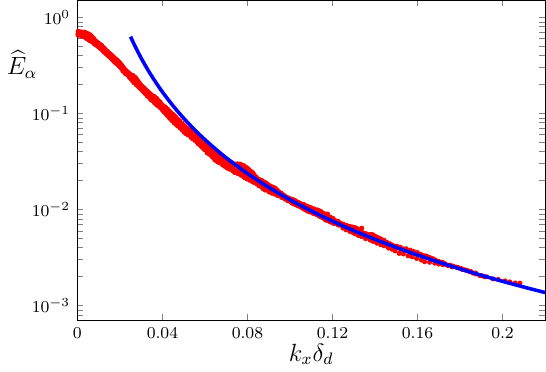}
   \includegraphics[width=0.49\textwidth]{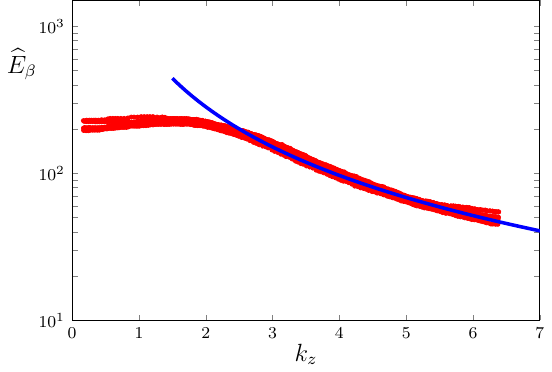}
   \put(-205,118){(a)}
   \put(-18,118){(b)}
  \end{picture}
  \caption{(a) Scaled energy spectrum ${\widehat E}_\alpha$ as a function of the streamwise wavenumber $k_x \delta_d$ and (b) scaled energy spectrum ${\widehat E}_\beta$ as a function of the spanwise wavenumber $k_z$. The experimental data by MA01, also shown in figures \ref{fig:matsubara-figure-13-14}(c,d), are represented by the red circles, and the algebraic best fitting lines in solid blue represent relations (a) \eqref{eq:e-alpha-fitting} and (b) \eqref{eq:e-beta-fitting}.}
  \label{fig:algebraic-fitting}
 \end{center}
\end{figure}

The data in figures \ref{fig:matsubara-figure-13-14}(c,d) are replotted in figure \ref{fig:algebraic-fitting}, which reveals that the experimental data of the energy spectra by MA01 are well approximated by the power laws

\begin{equation}
\label{eq:e-alpha-fitting}
\widehat E_\alpha = \frac{1.91\cdot 10^{-5}}{(k_x \delta_d)^{\widetilde \alpha}}, \quad \mbox{where} \quad \widetilde \alpha=2.82,
\end{equation}

\begin{equation}
\label{eq:e-beta-fitting}
\widehat E_\beta = \frac{8.3\cdot 10^2}{k_z^{\widetilde \beta}}, \quad \mbox{where} \quad \widetilde \beta=1.55.
\end{equation}
The power laws \eqref{eq:e-alpha-fitting} and \eqref{eq:e-beta-fitting} are useful in our theoretical analysis of \S\ref{sec:results}.

%------------------------------------------------------
\section{Theoretical framework for the Klebanoff modes}
\label{sec:flow-system}

The theory of the Klebanoff modes is found in LWG99. Here, we report the main points that are useful for our analysis of the wind-tunnel flow studied by MA01. 

\subsection{The free-stream disturbance flow at short streamwise distances}
\label{sec:free-stream}

A uniform flow of velocity $U_\infty^*$ past an infinitely thin flat plate transports homogeneous, statistically stationary vortical fluctuations of the gust type, i.e. disturbances that are convected passively by the mean flow. These free-stream perturbations are assumed to be of small amplitude with respect to $U_\iy^*$, so that the free-stream flow is represented as the sum of the mean uniform flow and the free-stream vortical disturbances, as

\begin{equation}
\label{eq:free-stream-perturbation}
 {\bf u}_\iy = {\bf \hat i} + \ep {\bf u'}_\iy (x - t, y , z) = 
 {\bf \hat i} + \ep \int_{-\iy}^\iy \int_{-\iy}^\iy \int_{-\iy}^\iy
 \hat{\bf u}'_\iy(k_x,k_y,k_z) \rme^{\mathrm{i}({\bf{k \cdot x}} - k_x t)} 
 \mathrm{d}k_x \mathrm{d}k_y \mathrm{d}k_z,
\end{equation}
where $\ep$$\ll$$1$, $\hat{\bf{u}}'_\iy$$=$$\{\hat u^\iy,\hat v^\iy,\hat w^\iy\}$$=$$\mathcal{O}(1)$, ${\bf k}$$=$$\{k_x,k_y,k_z\}$, and the streamwise wavenumber $k_x$ and the frequency $-k_x$ are related because of Taylor's hypothesis \citep{taylor-1938,hunt-1973}. In the experiments of MA01, the turbulence is generated by a grid located upstream of the leading edge of the plate, but we consider $x=0$ as the streamwise location where the free-stream turbulence starts influencing the system because that is where the turbulence intensity was measured by MA01, as explained in the second paragraph of p. 154 in MA01. The representation \eqref{eq:free-stream-perturbation} is valid at wall-normal distances that are sufficiently large for the flow not to be influenced by the presence of the boundary layer and the flat plate. The free-stream perturbation \eqref{eq:free-stream-perturbation} is not influenced by viscous dissipation while being transported downstream by the free-stream potential flow because it is valid only at sufficiently small $x$ location. The streamwise evolution of the free-stream flow is nevertheless taken into account at larger streamwise locations by the model of the free-stream spectrum studied in \S\S\ref{sec:variance} and \ref{sec:spectrum}, and by the numerical solution of the free-stream disturbance flow that includes the viscous dissipation and the inviscid displacement of the mean-flow streamlines due to the boundary layer, as discussed in \S\ref{sec:klebanoff-modes}. Furthermore, expansion \eqref{eq:free-stream-perturbation} is not valid for amplitudes of free-stream disturbances comparable with that of the mean flow and for a non-uniform free-stream mean flow because Taylor's hypothesis does not apply in those cases \citep{lundell-alfredsson-2004}.

%===============================
\subsection{The Klebanoff modes}
\label{sec:klebanoff-modes}

In the limit of large Reynolds number, $\R$$\gg$1, the mean laminar boundary layer that develops over the flat plate is described by the steady boundary-layer equations \citep{schlichting-gersten-2000}. The mean-flow streamwise and wall-normal velocity components are $U(x,y)$ and $V(x,y)$, and the wall-normal similarity coordinate is $\eta$=$y/\delta$=$y \sqrt{\R/2 x}$, where $\delta$=$\sqrt{2 x/\R}$=$\sqrt{2}\delta_d/1.72$ is the boundary-layer thickness used in LWG99.

The free-stream vortical flow encounters the boundary layer and generates the Klebanoff modes, as documented by the experimental data of MA01 discussed in \S\ref{sec:mat-alf-results}. We consider the limit $k_x$$=$$\mathcal{O}\left(\R^{-1}\right)$$\ll$$k_y,$$k_z$ because the Klebanoff modes are of low frequency. The boundary layer indeed acts as a low-frequency-pass filter and thus only the low-frequency disturbances penetrate into the boundary layer, as evidenced in figure 9b on p. 162 of MA01. 
We study the flow at downstream locations where $\delta^*$$=$$\mathcal{O}(\Lambda_z^*)$, and we scale the streamwise coordinate as $\xbar$$=$$k_x$$x$$=$$\mathcal{O}(1)$. As explained in LWG99, the condition for linearization in the boundary layer is $\ep/k_x$$\ll$$1$. The boundary-layer flow is expressed as the sum of the mean boundary-layer flow $\textbf{U}$ and the disturbance flow $\ep \textbf{u}'$, as follows (LWG99, \cite{hunt-1973}, \cite{hunt-carruthers-1990}):

\[
{\bf u} = {\bf U}(x,y) + \ep {\bf u'}(x,y,z,t) 
= {\bf U}(x,y) + \ep \int_{-\iy}^\iy \int_{-\iy}^\iy  
\hat{\bf{u}}'(x,y,k_x,k_z) \rme^{\mathrm{i}(k_z z - k_xt)} \mathrm{d}k_x \mathrm{d}k_z = 
\]

\begin{equation}
\label{eq:solution}
= \{U,V,0\} + \ep 
\int_{-\iy}^\iy \int_{-\iy}^\iy 
\left\{ \ubar_0 (\xbar,\eta), \left(\frac{2 \xbar k_x}{\R}\right)^{1/2} \vbar_0 (\xbar, \eta), \wbar_0 (\xbar,\eta) \right\} \rme^{\mathrm{i}(k_z z - k_xt)} \mathrm{d}k_x \mathrm{d}k_z + \mathcal{O}\left(\ep^2\right), 
\end{equation}
%================
\begin{sloppypar}
\noindent
where the leading-order velocity components with respect to $k_x$$\ll$$1$ are retained, i.e. 
$\{\ubar_0,\vbar_0,\wbar_0\}$=
$\left[\hat w^\iy+\mathrm{i} k_z \hat v^\iy/\left(k_x^2+k_z^2\right)^{1/2}\right]$
$\left\{(\mathrm{i} k_z/k_x) \ubar,\left(\mathrm{i} k_z/k_x\right) \vbar,\wbar\right\}$. 
The components $\{\ubar,\vbar,\wbar\}$ satisfy the linearized unsteady boundary-region equations, complemented by initial and boundary conditions, all found in LWG99. Homogeneous boundary conditions at the wall represent the no-slip condition, while mixed boundary conditions in the free stream account for the boundary-layer inviscid displacement and the perturbation decay due to viscous dissipation. The system is solved by a second-order implicit finite-difference scheme and a standard block-elimination algorithm \citep{ricco-wu-2007}, described in Appendix \ref{sec:num-meth}.
\end{sloppypar}

The scaled wavenumber $\kz$$=$$k_z/(k_x \R)^{1/2}$$=$$\mathcal{O}(1)$ represents the relative importance between spanwise and wall-normal viscous effects at $\overline{x}$$=$$\mathcal{O}(1)$. In the limit $\kz$$\ll$$1$, the spanwise viscous diffusivity becomes negligible and the dynamics is ruled by the boundary-layer equations.

\begin{figure}
 \begin{center}
  \includegraphics[width=0.75\textwidth]{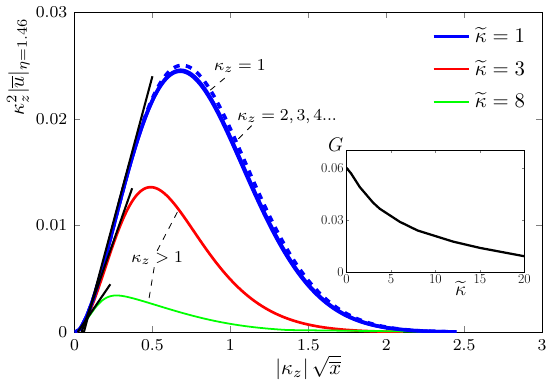}
  \caption{Growth and decay of the scaled streamwise velocity component of the Klebanoff modes $|\ut|=\kz^2 |\overline{u}|$ at $\eta=1.46$ as a function of the scaled streamwise coordinate $\xt=\abs{\kz} \sqrt{\overline{x}}$ for different $\kt$ values. The velocity is computed by solving numerically the boundary-region equations, found in LWG99. The straight solid lines denote the linear growth. The inset shows the slope of the linear growth, $G(\kt)$.}
  \label{fig:u-versus-sqrt-x}
 \end{center}
\end{figure}

We now discuss an asymptotic result, based on the parameter $\kz$, which is central in the analysis developed in \S\ref{sec:results}.
LWG99 showed that an asymptotic solution exists in the low-frequency, large-spanwise-wavenumber limit $\kz$$\gg$1 with $\kt$$=$$\ky/\abs{\kz}$$=$$\mathcal{O}$$(1)$, where $\ky$$=$$k_y/(k_x \R)^{1/2}$. In this limit, the leading-order velocity components $\{\ubar,\vbar,\wbar\}$ are rescaled and expressed as a function of the new streamwise coordinate $\xt$=$\kz^2 \xbar$$=$$\mathcal{O}(1)$, i.e. $\ut(\xt,\eta,\kt)$=$\kz^2 \ubar$$=$$\mathcal{O}(1)$, $\{\vt,\wt\}(\xt,\eta,\kt)$=$\{\vbar,\wbar\}$$=$$\mathcal{O}(1)$. The rescaled velocity components $\{\ut,\vt,\wt\}$ are quasi-steady and depend only on the ratio of wavenumbers $\kt$ and not explicitly on the scaled spanwise wavenumber $\kz$. Although the asymptotic solution is valid for $\kz$$\gg$1, the numerical calculations reveal the remarkable result that the algebraic growth of the quasi-steady asymptotic solution $\{\ut,\vt,\wt\}$ is indistinguishable from the full boundary-region solution even for $\kz$ as low as 1. 
Figure \ref{fig:u-versus-sqrt-x} indeed shows that the trends of $\abs{\ut}$ for different $\kz$$\geq$$1$ and the same $\kt$ collapse onto one another when plotted as a function of $\xt$. It also means that the asymptotic solution describes the Klebanoff modes well even when the spanwise wavelength is comparable with the boundary-layer thickness, which is precisely the flow condition of interest in the experiments of MA01. Therefore, the asymptotic solution $\{\ut,\vt,\wt\}$ is utilized in the scaling analysis of \S\ref{sec:results}, where the collapse of the spectral distributions shown in figure \ref{fig:algebraic-fitting} is obtained. Figure \ref{fig:u-versus-sqrt-x} also reveals that the initial growth of the disturbance is linear when $\kz \geq 1$, that is, $|\ut|= G(\kt)\abs{\kz} \sqrt{\overline{x}}$. 
The inset of figure \ref{fig:u-versus-sqrt-x} shows the slope $G(\kt)$. The decay of $\ut$ as $\kt$$\rightarrow$$\infty$, and therefore of $G(\kt)$, is predicted by the asymptotic analysis because, in the limits $\kappa_z$$\gg$$1$ and $\kt$$\gg$$1$, the solution can be written as $\ub(\xb,\eta)={\kt}^2 \ut=\ky^2 \ubar=\mathcal{O}(1)$, where $\xb={\kt}^2 \xt=\ky^2 \xbar$. 

%---------------------------------------
\section{Scaling of the Klebanoff modes}
\label{sec:results}
%--------------------------------------------------------------
\subsection{Variance of the boundary-layer streamwise velocity}
\label{sec:variance}

The boundary-layer perturbations and the free-stream modes are related as \citep{hunt-1973,hunt-carruthers-1990}

\[
\hat{u}_i'(x,y,k_x,k_z) 
= \int_{-\iy}^\iy M_{ij}(x,y;k_x,k_y,k_z) \hat{u}_{\iy j}'(k_x,k_y,k_z) \mathrm{d}k_y,
\]
where $M_{ij}$ is a tensor acting as a transfer function between the free-stream flow and the boundary-layer flow.
The interest is in the correlation of the boundary-layer velocity components, delayed in time and $z$ \citep{batchelor-1953},

\[
R_{ij}(x,y,r_z,\tau) = \ep^2 \averzt{u^\prime_i(x,y,z+r_z,t+\tau)u^\prime_j(x,y,z,t)},
\]
which can be expressed as (refer to pp. 638-640 in \cite{hunt-1973})

\[
R_{ij}(x,y,r_z,\tau) = \ep^2 \int_{-\iy}^\iy \int_{-\iy}^\iy \int_{-\iy}^\iy 
\sum_{l=1}^3 \sum_{m=1}^3 M_{il}^\dagger M_{jm} \Phi_{\iy lm}({\bf k}) 
\rme^{\mathrm{i}(k_z r_z - k_x \tau)}
\mathrm{d}k_x \mathrm{d}k_y \mathrm{d}k_z,
\]
where $\Phi_{\iy lm}$ is the spectral tensor of the turbulence upstream of the flat plate and the symbol $\dagger$ indicates the complex conjugate. The focus is on the spectral properties of the mean-square streamwise velocity fluctuations, i.e. $i$$=$$j$$=$$1$, $r_z$$=$$\tau$$=$$0$ (LWG99),

\begin{equation}
\label{eq:r.m.s.}
\ep^2 \averzt{u^{\prime 2}} = R_{11}(x,y,0,0) = \ep^2 \int_{-\iy}^\iy \int_{-\iy}^\iy \int_{-\iy}^\iy 
\sum_{l=1}^3 \sum_{m=1}^3 M_{1l}^\dagger M_{1m}\Phi_{\iy lm}({\bf k}) \mathrm{d}k_x \mathrm{d}k_y \mathrm{d}k_z.
\end{equation}
The relevant components of the transfer-function tensor $M_{ij}$ are

\begin{equation}
\label{eq:M} 
M_{11} = \ubar^{(0)}, \quad
M_{12} = \frac{\mathrm{i} k_x}{\sqrt{k_x^2+k_z^2}} \ubar^{(0)} - \frac{k_z^2}{k_x \sqrt{k_x^2+k_z^2}} \ubar, \quad
M_{13} = \frac{\mathrm{i} k_z}{k_x} \ubar,
\end{equation}
where $\ubar^{(0)}$ is the next-order term of the expansion of $\overline{u}_0$ in \eqref{eq:solution} with respect to $k_x$$\ll$$1$ (LWG99). By substituting \eqref{eq:M} into \eqref{eq:r.m.s.} and collecting the dominant ter.m.s. $\mathcal{O}\left(k_x^{-2}\right)$, the integrand in \eqref{eq:r.m.s.} becomes

\begin{equation}
\label{eq:integrand} 
\sum_{l=1}^3 \sum_{m=1}^3 M_{1l}^\dagger M_{1m}\Phi_{\iy lm}({\bf k}) = 
\frac{k_z^2 \abs{\ubar}^2}{k_x^2}
\left(
\frac{k_z^2}{\sqrt{k_x^2+k_z^2}} \Phi_{\iy 22} +
\Phi_{\iy 33}
\right) 
+ \mathcal{O}(k_x^{-1}).
\end{equation}
As suggested by LWG99 on p. 187, an axial-symmetric turbulence model that describes free-stream turbulence is \citep{batchelor-1953,chandrasekhar-1950}

\begin{equation}
\label{eq:spectrum}
\Phi_{\iy ij} = \frac{ k_\perp^2 \delta_{ij}^\perp - k_{\perp i} k_{\perp j} }{k_\perp^2} 
\left( \Phi_t - \frac{2 k_x^2}{k_\perp^2}  \Phi_x \right) + \frac{\Phi_x}{k_\perp^2}
\left( k_x^2 \delta_{ij}^\perp - k_x k_{\perp i} \delta_{i1} + k_\perp^2 \delta_{i1} \delta_{j1} \right),
\end{equation}
where $k_{\perp i} = k_i - \delta_{i1} k_x$, $\delta_{i1}$ is the Kronecker delta, $\delta_{ij}^\perp = \delta_{ij} - \delta_{i1} \delta_{j1}$ is the cross-stream Kronecker delta, and $k_\perp$$=$$\sqrt{k_y^2+k_z^2}$. The functions $\Phi_x$$=$$\Phi_x(k_x,k_\perp)$ and $\Phi_t$$=$$\Phi_t(k_x,k_\perp)$ are the longitudinal and transverse spectra. In the limit $k_x$$\rightarrow$0,

\begin{equation}
\label{eq:phi-22-33} 
\Phi_{\iy 22} = \frac{k_z^2}{k_x^2+k_z^2}\Phi_t, \quad \Phi_{\iy 33} = \frac{k_x^2}{k_x^2+k_z^2} \Phi_t.
\end{equation}
Substitution of \eqref{eq:phi-22-33} into \eqref{eq:integrand} and then into \eqref{eq:r.m.s.} leads to the variance of boundary-layer streamwise velocity,

\begin{equation}
\label{eq:r.m.s.-simple}
\averzt{u^{\prime 2}}(x,y) =  \int_{-\iy}^\iy \int_{-\iy}^\iy \int_{-\iy}^\iy
\left(\frac{k_z}{k_x}\right)^2 \abs{\ubar}^2(x,y) \Phi_t(k_x,k_\perp) \mathrm{d}k_x \mathrm{d}k_y \mathrm{d}k_z.
\end{equation}
As discussed by LWG99, these results demonstrate that, at leading order, the growth and development of the Klebanoff modes is dictated by the transverse spectral function $\Phi_t$ obtained by correlations of the velocity components perpendicular to the streamwise direction (refer to LWG99 on p. 188), and not by the longitudinal spectral function $\Phi_x$, which is typically the object of experimental investigations of freely decaying grid-generated turbulence. 

%-------------------------------------------
\subsection{Free-stream turbulence spectrum}
\label{sec:spectrum}

The axial-symmetric transverse turbulence spectrum $\Phi_t(k_x,k_\perp)$ in \S\ref{sec:variance} is assumed to pertain to homogeneous turbulence and it is therefore independent of the streamwise direction \citep{hunt-1973,hunt-carruthers-1990}. However, in a more general non-homogeneous case, the turbulence spectrum also depends on the position vector, $\Phi_t(\textbf{x},k_x,k_\perp)$, as for example discussed in \cite{townsend-1980}. To the best of our knowledge, no detailed measurements of $\Phi_t$ have been made, so our objective is to suggest a functional form for $\Phi_t$ that is a satisfactory model for our problem. 

Our choice of spectrum takes inspiration from the theory of temporally decaying turbulence discussed in \cite{townsend-1980} on p. 61. The results in the streamwise decaying case can be assumed to be qualitatively analogous to the temporally decaying case if the streamwise direction is considered in lieu of time for flows where the turbulence intensity is much smaller than the free-stream mean velocity, i.e. when Taylor's hypothesis is valid, as explained in \cite{townsend-1980} on p. 65.
In the idealized limit of vanishingly small amplitude of free-stream turbulence generated by a grid swept through a still fluid, \cite{batchelor-1953}, on p. 93, shows that the time dependency is due solely to the viscous dissipation, and the temporal decay is exponential. However, \cite{batchelor-1953} warns that this behaviour would occur only after a long time and it would not apply to a real turbulent flow generated by a grid in a wind tunnel. The exponential decay would thus not pertain to locations relatively close to the turbulence-generating grid, which are certainly of interest in the study of the MA01 experimental results. Furthermore, if the turbulence spectrum $\Phi_t$ were assumed to be independent of the streamwise direction, as in \S\ref{sec:variance}, the streamwise evolution of the free-stream disturbance would affect the variance $\averzt{u^{\prime 2}}$ in the boundary layer only indirectly through the decaying free-stream wall-normal and spanwise velocity components because $\abs{\ubar}$, the leading-order component in \eqref{eq:r.m.s.-simple}, vanishes as $y$$\rightarrow$$\infty$ (refer to equations (5.11) and (5.20)-(5.22) in LWG99). Neglecting the streamwise dependency of the free-stream spectrum would mean that the free-stream decay would be purely exponential because it is dictated by a linearized dynamics.
Including the streamwise dependence in $\Phi_t$ is therefore deemed to be more realistic and it also serves the purpose of modelling mild effects of nonlinearity. Similar modelling of mild nonlinearity in a free-stream spectrum pertaining to realistic grid-generating turbulence has been proposed by LWG99 in their \S7.2.

\cite{townsend-1980} on p. 61 shows that the spectral function for decaying turbulence has the form

\begin{equation}
\label{eq:isotropic-spectrum}
E(k,t) = \averyz{u'(t)^2} L(t) \mathcal{F}(k L(t)),
\end{equation}
where $L(t)$ is an integral scale representing the free-stream isotropic turbulence, $\averyz{\cdot}$ indicates spatial averaging and $k$ is the wavenumber. The spectral function \eqref{eq:isotropic-spectrum} is found by appropriate scaling of experimental data \citep{stewart-townsend-1951}, as also discussed in \cite{hinze-1975} on p. 263. By substitution of \eqref{eq:isotropic-spectrum} into the equation governing the rate of change of the turbulence spectrum, \cite{townsend-1980} finds

\begin{equation}
\label{eq:space-time-scales}
\frac{\mathrm{d} \averyz{u'(t)^2}}{\mathrm{d} t} \propto  \frac{ \averyz{u'(t)^2}^{3/2}}{L(t)}, \quad
\frac{\mathrm{d} L(t)}{\mathrm{d} t} \propto \averyz{u'(t)^2}^{1/2},
\end{equation}
as further explained in \cite{batchelor-1953} on p. 103. The temporal decay of $\averyz{u'(t)^2}$ that satisfies \eqref{eq:space-time-scales} is 

\begin{equation}
\label{eq:energy-time-decay}
\averyz{u'(t)^2} \propto t^{-\gamma},
\end{equation}
which is consistent with numerous experimental data, for which 1.15$<$$\gamma$$<$1.45 (refer to p. 160 of \cite{pope-2000}), and with theoretical studies, which suggest $\gamma$$=$1 (refer to \cite{tennekes-lumley-1972}) or $\gamma$$=$3/2 (refer to \cite{davidson-2004} on p. 407, where the Saffman spectrum is discussed). The decay constant $\gamma$ can then be assumed to be 

\begin{equation}
\label{eq:alpha-range}
1 \leq \gamma \leq 3/2.
\end{equation}
The integral spatial scale $L$ is predicted to grow as

\begin{equation}
\label{eq:L}
L \propto t^\zeta, \quad 1/4 \leq \zeta \leq 1/2.
\end{equation}
Substitution of \eqref{eq:energy-time-decay} and \eqref{eq:L} into \eqref{eq:isotropic-spectrum}, and use of \eqref{eq:alpha-range}, lead to a simplified form of the spectrum

\begin{equation}
\label{eq:isotropic-spectrum-simple}
E(k,t) \propto \frac{\mathcal{F}\left(k t^{d/2}\right)}{t^{c/2}}, 
\end{equation}
for which the inequalities 

\begin{equation}
\label{eq:inequalities}
1 \leq c \leq 5/2 \quad \mbox{and} \quad 1/2 \leq d \leq 1
\end{equation}
apply. 
%\fcolorbox{black}{cyan}{$1 \leq c \leq 5/2$}, 
%\fcolorbox{black}{cyan}{$1/2 \leq d \leq 1$}. 
Also, $c$$=$$3\gamma-2$ and $d=2-\gamma$, from which 

\begin{equation}
\label{eq:equality}
c=4-3d.
\end{equation}
%\fcolorbox{black}{cyan}{$c=4-3d$}. 

The time-decaying isotropic spectrum \eqref{eq:isotropic-spectrum-simple} can now be used to obtain a spectrum that pertains to the grid-generated turbulence of interest in our problem. As the spectrum has to account for the streamwise decay of turbulence, the temporal dependence in \eqref{eq:isotropic-spectrum-simple} is converted to the streamwise dependence. The axial symmetry of the turbulence has to be modelled by including the effect of the cross-flow wavenumber $k_\perp$ because, as explained by \cite{batchelor-1953}, purely isotropic turbulence is extremely hard to obtain in the laboratory. 
Our axial-symmetric transverse spectrum therefore reads:

\begin{equation}
\label{eq:phi-t}
\Phi_t(x,k_x,k_\perp;\R) = \frac{1}{k_\perp^b (k_x \R)^2 \delta^c} \mathcal{F}\left(\frac{k_x \R \delta^d}{k_\perp^n}\right),
\end{equation}
where, in lieu of $t$ in \eqref{eq:isotropic-spectrum-simple}, we have introduced the streamwise coordinate $x$ and expressed this dependence through the boundary-layer thickness $\delta$ because $\delta \propto \sqrt{x}$. The dependence of the spectrum on $k_\perp$ is introduced inside and outside the function $\mathcal{F}$ to allow maximum generality. The spatial dependence of the spectrum \eqref{eq:phi-t} is mild compared with the long streamwise length scale of the Klebanoff modes because \eqref{eq:phi-t} is  expressed as a function of $\delta=\delta^*/\Lambda_z^*$, where $\delta^*$ and $\Lambda_z^*$ are comparable.
Consistently with the theoretical framework of \S\ref{sec:flow-system}, the low-frequency assumption is adopted as the boundary layer acts as a low-frequency-pass filter. It is thus reasonable to consider a free-stream spectrum such as \eqref{eq:phi-t}, dominated by low-frequency disturbances ($k_x\ll1$ with $k_x R_\lambda=\mathcal{O}(1)$ or smaller).

The parameters $n,b,c,d$ in \eqref{eq:phi-t} are found by asymptotic analysis and by fitting the experimental data. The parameters $c$ and $d$ play analogous roles in \eqref{eq:isotropic-spectrum-simple} and \eqref{eq:phi-t}. 

%-----------------------------------------------------------------
\subsection{Scaling of boundary-layer streamwise velocity spectra}

By substituting the spectrum $\eqref{eq:phi-t}$ into \eqref{eq:r.m.s.-simple}, the variance of the boundary-layer streamwise velocity becomes

%---------------
\begin{equation}
\label{eq:variance-u}
\averzt{u^{\prime 2}}
= \int_{-\iy}^\iy \int_{-\iy}^\iy \int_{-\iy}^\iy 
\left(\frac{k_z}{k_x}\right)^2\frac{\abs{\overline{u}}^2}{k_\perp^b (k_x \R)^2 \delta^c} \mathcal{F}\left(\frac{k_x \R \delta^d}{k_\perp^n}\right) \mathrm{d}k_x \mathrm{d}k_y \mathrm{d}k_z.
\end{equation}

Expression \eqref{eq:variance-u} is used with \eqref{eq:spectra} and \eqref{eq:ealpha-ebeta-hats} to explain the scaling of the experimental results, shown in figures \ref{fig:matsubara-figure-13-14}(c,d). The four parameters $n,b,c,d$ in \eqref{eq:variance-u} are found by using the following four conditions.

\begin{itemize}
    \item In figure \ref{fig:matsubara-figure-13-14}(c), the spectrum $\widehat E_\alpha$ depends only on the scaled streamwise wavenumber $k_x \delta_d$.
    \item In figure \ref{fig:matsubara-figure-13-14}(d), the spectrum $\widehat E_\beta$ depends only on the spanwise wavenumber $k_z$ and is independent of the streamwise location.
    \item In figure \ref{fig:algebraic-fitting}(a), the best fitting of the experimental data leads to the power-law dependency \eqref{eq:e-alpha-fitting} for $\widehat E_\alpha(k_x \delta_d)$. 
    \item In figure \ref{fig:algebraic-fitting}(a), the best fitting of the experimental data leads to the power-law dependency \eqref{eq:e-beta-fitting} for $\widehat E_\beta(k_z)$.
\end{itemize}

%==================================================
\subsubsection{Spectrum versus spanwise wavenumber}
\label{sec:e-beta}

Motivated by the scaling of the spectrum $E_\beta$ by $Re_x$, given in the second expression in \eqref{eq:ealpha-ebeta-hats}, the variance \eqref{eq:variance-u} is rescaled by $Re_x$ as

%---------------
\begin{equation}
\frac{\averzt{u^{\prime 2}}}{C_e Re_x} 
= \int_{-\iy}^\iy \int_{-\iy}^\iy \int_{-\iy}^\iy 
\frac{k_z^2 \abs{\overline{u}}^2}{C_e k_\perp^b (k_x \R)^3 \overline{x} \delta^c} \mathcal{F}\left(\frac{k_x \R \delta^d}{k_\perp^n}\right) \mathrm{d}k_x \mathrm{d}k_y \mathrm{d}k_z.
\end{equation}
The streamwise velocity $\abs{\overline{u}}$ is changed to $\abs{\overline{u}}^2= \abs{\widetilde{u}}^2 k_x^2 \R^2/k_z^4$ and the streamwise coordinate is eliminated by using $\overline{x}=\delta^2 k_x \R/2$, to obtain
\begin{comment}
%---------------
\begin{equation}
\frac{\averzt{u^{\prime 2}}}{C_e Re_x} 
= \int_{-\iy}^\iy \int_{-\iy}^\iy \int_{-\iy}^\iy 
\frac{\abs{\widetilde{u}}^2}{C_e k_z^2 k_\perp^b k_x \R \overline{x} \delta^c} \mathcal{F}\left(\frac{k_x \R \delta^d}{k_\perp^n}\right) \mathrm{d}k_x \mathrm{d}k_y \mathrm{d}k_z.
\end{equation}
\end{comment}

%---------------
\begin{equation}
\label{eq:u-prime-middle2}
\frac{\averzt{u^{\prime 2}}}{C_e Re_x} 
= \int_{-\iy}^\iy \int_{-\iy}^\iy \int_{-\iy}^\iy 
\frac{2 \abs{\widetilde{u}}^2}{C_e k_z^2 k_\perp^b (k_x \R)^2 \delta^{c+2}} \mathcal{F}\left(\frac{k_x \R \delta^d}{k_\perp^n}\right) \mathrm{d}k_x \mathrm{d}k_y \mathrm{d}k_z.
\end{equation}
The asymptotic solution for $\kz \gg 1$, i.e. $\abs{\widetilde{u}}^2 = (k_z \delta)^2 \abs{G(\kt)}^2/2$, shown in figure \ref{fig:u-versus-sqrt-x} and discussed at the end of \S\ref{sec:klebanoff-modes}, is substituted into \eqref{eq:u-prime-middle2} to arrive at

%---------------
\begin{equation}
\frac{\averzt{u^{\prime 2}}}{C_e Re_x} 
= \int_{-\iy}^\iy \int_{-\iy}^\iy \int_{-\iy}^\iy 
\frac{\abs{G(\kt)}^2}{C_e k_\perp^b (k_x \R)^2 \delta^c} \mathcal{F}\left(\frac{k_x \R \delta^d}{k_\perp^n}\right) \mathrm{d}k_x \mathrm{d}k_y \mathrm{d}k_z.
\end{equation}
The wavenumbers $k_\perp$ and $k_y$ are eliminated by using $k_\perp$=$|k_z| (1+\kt^2)^{1/2}$=$|k_z| K(\kt)$ and $k_y = k_z \kt$, and the integration limits are changed to [0,$\infty$)

%---------------
\begin{equation}
\frac{\averzt{u^{\prime 2}}}{C_e Re_x} 
= \int_0^\iy \int_0^\iy \int_0^\iy 
\frac{2^3 \abs{G(\kt)}^2}{C_e k_z^{b-1}(k_x \R)^2 K(\kt)^b \delta^c} 
\mathcal{F}\left(\frac{k_x \R \delta^d}{(k_z K(\kt))^n}\right) \mathrm{d}k_x \mathrm{d}\kt \mathrm{d}k_z.
\end{equation}
By using the rescaled \eqref{eq:spectra}, 

%---------------
\begin{equation}
\frac{\averzt{u^{\prime 2}}}{C_e Re_x} 
= \frac{C_\beta}{\epsilon^2} \int_0^\iy \widehat E_\beta (k_z) \mathrm{d}k_z,
\end{equation}
%--------------
\begin{comment}
%---------------
\begin{equation}
\widehat E_\beta (k_z)
= 
\frac{\epsilon^2}{C_e C_\beta}
\int_0^\iy \int_0^\iy 
\frac{2^3 \abs{G(\kt)}^2}{k_z^{b-1}(k_x \R)^2 K(\kt)^b \delta^c} 
\mathcal{F}\left(\frac{k_x \R \delta^d}{(k_z K(\kt))^n}\right) \mathrm{d}k_x \mathrm{d}\kt.
\end{equation}
\end{comment}
we find

%---------------
\begin{equation}
\label{eq:e-beta-rearranged}
\widehat E_\beta (k_z)
= 
\frac{2^3 \epsilon^2}{C_e C_\beta \R^2 k_z^{b-1} \delta^c} 
\int_0^\iy 
\frac{\abs{G(\kt)}^2}{K(\kt)^b}
\underbrace{
\int_0^\iy
    \frac{1}{k_x^2}\mathcal{F}\left(\frac{k_x \R \delta^d}{(k_z K(\kt))^n}\right) 
\mathrm{d}k_x
}_{I_\beta}
\mathrm{d}\kt.
\end{equation}
By defining the integration variable $\sigma = k_x \R \delta^d/\left[k_z K(\kt)\right]^n$, the integral $I_\beta$ in \eqref{eq:e-beta-rearranged} becomes

\begin{equation}
\label{eq:ib}
I_\beta = 
\frac{\R \delta^d}{[k_z K(\kt)]^n }
\int_0^\iy \frac{\mathcal{F}(\sigma)}{\sigma^2} \mathrm{d}\sigma. 
\end{equation}
Upon substitution of \eqref{eq:ib} into \eqref{eq:e-beta-rearranged}, we obtain

\begin{equation}
\label{eq:e-beta-final}
\widehat E_\beta (k_z)
= 
\frac{2^3 \epsilon^2 \delta^{d-c}}{C_e C_\beta \R k_z^{b-1+n}} 
\int_0^\iy \frac{\abs{G(\kt)}^2}{[K(\kt)]^{n+b}} \mathrm{d}\kt
\int_0^\iy \frac{\mathcal{F}(\sigma)}{\sigma^2} \mathrm{d}\sigma.
\end{equation}
The key point here is that, as the function $\widehat E_\beta$ must not depend on the streamwise direction, the dependence on $\delta$ must be eliminated. It follows that $c=d$. %\fcolorbox{black}{yellow}{$c=d$}. 

The spectrum \eqref{eq:e-beta-final} becomes

\begin{equation}
\label{eq:algebraic-beta}
\widehat E_\beta (k_z)
= \frac{B_\beta G_\beta \Sigma_\beta}{{k_z}^{\widetilde \beta}},
\end{equation}
where $\widetilde \beta = b -1 + n$, and
%\fcolorbox{black}{yellow}{$\widetilde \beta = b -1 + n$},

%---------------
\begin{equation}
B_\beta = \frac{2^3 \epsilon^2}{\R C_e C_\beta}, \quad 
G_\beta = 
\int_0^\iy \frac{\abs{G(\kt)}^2}{\left(1+\kt^2\right)^{(n+b)/2}} \mathrm{d}\kt, \quad
\Sigma_\beta = \int_0^\iy \frac{\mathcal{F}(\sigma)}{\sigma^2} \mathrm{d}\sigma.
\end{equation}
The algebraic decay emerging in \eqref{eq:algebraic-beta} matches the behaviour of the experimental data in figure \ref{fig:algebraic-fitting} (right). At small $k_z$, the theoretical framework does not predict the trend of the data in figure \ref{fig:algebraic-fitting} (right), which is almost independent of $k_z$. At small $k_z$, the spanwise wavelength is larger than the boundary-layer thickness, the spanwise viscous effects are negligible and the flow is ruled by the boundary-layer equations, as discussed in \S\ref{sec:free-stream}. Our analysis instead hinges on the asymptotic solution of the boundary-region equations for which the spanwise wavelength and the boundary-layer thickness are comparable, i.e. the wall-normal and spanwise diffusion effects are both important ($\kappa_z=\mathcal{O}(1)$ or larger). The same reasoning applies to the dash-dotted lines in figure \ref{fig:matsubara-figure-13-14}(d), which do not collapse onto one another as they correspond to streamwise locations close to the leading edge, where spanwise-diffusion effects are negligible.

%====================================================
\subsubsection{Spectrum versus streamwise wavenumber}
\label{sec:e-alpha}

Motivated by the scaling of the spectrum $E_\alpha$ by $Re_x^{3/2}$, given in the first expression in \eqref{eq:ealpha-ebeta-hats}, the variance \eqref{eq:variance-u} is rescaled by $Re_x^{3/2}$. By using $c=d$, found in \S\ref{sec:e-beta}, we find

%---------------
\begin{equation}
\frac{\averzt{u^{\prime 2}}}{C_e Re_x^{3/2}} 
= 
\int_{-\iy}^\iy \int_{-\iy}^\iy \int_{-\iy}^\iy 
\frac{k_z^2 \abs{\overline{u}}^2}{C_e k_x^{5/2} k_\perp^b \R^{7/2} \delta^c \overline{x}^{3/2}} 
\mathcal{F}\left(\frac{k_x \R \delta^c}{k_\perp^n}\right) \mathrm{d}k_x \mathrm{d}k_y \mathrm{d}k_z.
\end{equation}
The streamwise velocity $\abs{\overline{u}}$ is changed to $\abs{\overline{u}}^2= \abs{\widetilde{u}}^2 k_x^2 \R^2/k_z^4$ and the streamwise coordinate is eliminated by using $\overline{x}=\delta^2 k_x \R/2$ to find

\begin{comment}
%---------------
\begin{equation}
\frac{\averzt{u^{\prime 2}}}{C_e Re_x^{3/2}}
= \int_{-\iy}^\iy \int_{-\iy}^\iy \int_{-\iy}^\iy 
\frac{\abs{\widetilde u}^2}{C_e k_z^2 k_x^{1/2} k_\perp^b \R^{3/2} \delta^c \overline{x}^{3/2}} 
\mathcal{F}\left(\frac{k_x \R \delta^c}{k_\perp^n}\right) \mathrm{d}k_x \mathrm{d}k_y \mathrm{d}k_z.
\end{equation}
\end{comment}
%---------------
\begin{equation}
\label{eq:e-alpha-middle}
\frac{\averzt{u^{\prime 2}}}{C_e Re_x^{3/2}}
= \int_{-\iy}^\iy \int_{-\iy}^\iy \int_{-\iy}^\iy 
\frac{2^{3/2}\abs{\widetilde u}^2}{C_e k_z^2 k_x^2 k_\perp^b \R^3 \delta^{c+3}} 
\mathcal{F}\left(\frac{k_x \R \delta^c}{k_\perp^n}\right) \mathrm{d}k_x \mathrm{d}k_y \mathrm{d}k_z.
\end{equation}
The changes of variable $k_z = k_\perp \sin \theta$, $k_y = k_\perp \cos \theta$, 
$k_\perp =k_o/\delta$ are used in \eqref{eq:e-alpha-middle} to find

%---------------
\begin{equation}
\label{eq:u-prime-middle}
\frac{\averzt{u^{\prime 2}}}{C_e Re_x^{3/2}}
= \int_{-\iy}^\iy \int_0^\iy \int_0^{2 \pi} 
\frac{2^{3/2}\abs{\widetilde u}^2}{C_e \delta^{c+3-b} k_o^{b+1} (\sin \theta)^2 k_x^2 \R^3} 
\mathcal{F}\left(\frac{k_x \R \delta^{n+c}}{k_o^n}\right) \mathrm{d}\theta \mathrm{d}k_o \mathrm{d}k_x.
\end{equation}
We substitute the asymptotic result 

\[
\abs{\widetilde u}^2 
= 
\frac{k_z^2 \delta^2}{2} \abs{G(\widetilde \kappa)}^2 
=
\frac{k_o^2 (\sin \theta)^2}{2} \abs{G(\cot \theta)}^2
\]
into \eqref{eq:u-prime-middle} to obtain
%---------------
\begin{equation}
\label{eq:ea}
\frac{\averzt{u^{\prime 2}}}{C_e Re_x^{3/2}}
= \int_{-\iy}^\iy \int_0^\iy \int_0^{2 \pi} 
\frac{\sqrt{2} \abs{G(\cot \theta)}^2 }{C_e \delta^{c+4-b} k_o^{b-1} k_x^2 \R^3} 
\mathcal{F}\left(\frac{k_x \R \delta^{n+c}}{k_o^n}\right) \mathrm{d}\theta \mathrm{d}k_o \mathrm{d}k_x.
\end{equation}
By using the rescaled \eqref{eq:spectra}, 

%---------------
\begin{equation}
\label{eq:sp-ealpha}
\frac{\averzt{u^{\prime 2}}}{C_e Re_x^{3/2}} 
= \frac{C_\alpha}{\epsilon^2} \int_0^\iy \frac{\widehat E_\alpha (k_x \delta)}{\delta} \mathrm{d}(k_x \delta),
\end{equation}
changing the limits of the integration along $k_x$ to $[0,\infty)$, and equating \eqref{eq:ea} and \eqref{eq:sp-ealpha}, we find
\begin{comment}
\begin{equation}
\label{eq:sp-ealpha-2}
\widehat E_\alpha (k_x \delta)=  
\frac{2^{3/2} \epsilon^2}{C_e C_\alpha \R^3} 
\int_0^{2 \pi} \abs{G(\cot \theta)}^2 \mathrm{d}\theta
\int_0^\iy \mathcal{F}\left(\frac{k_x \R \delta^{n+c}}{k_o^n}\right) \frac{\mathrm{d}k_o}{k_x^2 \delta^{c+3-b} k_o^{b-1}}.
\end{equation}
\end{comment}

\begin{equation}
\label{eq:sp-ealpha-3}
\widehat E_\alpha (k_x \delta)=  
\frac{2^{3/2} \epsilon^2 G_\alpha}{C_e C_\alpha \R^3}
\int_0^\iy \mathcal{F}\left(\frac{k_x \R \delta^{n+c}}{k_o^n}\right) \frac{\mathrm{d}k_o}{k_x^2 \delta^{c+3-b} k_o^{b-1}},
\end{equation}
where

\begin{equation}
G_\alpha = \int_0^{2 \pi} \abs{G(\cot \theta)}^2 \mathrm{d}\theta.
\end{equation}
We define the integration variable $\omega = k_x \R \delta^{n+c}/k_o^n$ ($n>0$) in \eqref{eq:sp-ealpha-3} to obtain

\begin{comment}
It follows that
\begin{equation}
\mathrm{d}k_o = -\left(\frac{k_x \R \delta^{c+n}}{\omega}\right)^{\frac{n+1}{n}} \frac{\omega}{n k_x \R \delta^{c+n}}.
\end{equation}
\begin{equation}
k_o^{b-1} = \left(\frac{k_x \R \delta^{c+n}}{\omega}\right)^{\frac{b-1}{n}}.
\end{equation}
\end{comment}
\begin{align}
\label{eq:sp-ealpha-3b}
\widehat E_\alpha (k_x \delta)=  
\frac{2^{3/2} \epsilon^2 G_\alpha}{n C_e C_\alpha \R^{3+(b-2)/n}}
\int_0^\iy \frac{\mathcal{F}\left( \omega \right)\mathrm{d}\omega}{\omega^{1+(2-b)/n} \delta^{c+1+c(b-2)/n} k_x^{2+(b-2)/n}}.
\end{align}
By defining $\widetilde \alpha = 2+(b-2)/n$ and $\widetilde d = c + 1 + c(b-2)/n$, the spectrum becomes

\begin{align}
\label{eq:sp-ealpha-4}
\widehat E_\alpha (k_x \delta)
%&
%\frac{2^{3/2} \epsilon^2 G_\alpha}{n C_e C_\alpha \R^{3-(2-b)/n}}
%\frac{1}{\delta^{\widetilde d} k_x^{\widetilde \alpha}}
%\int_0^\iy \frac{\mathcal{F}\left( \omega \right)\mathrm{d}\omega}{\omega^{1+(2-b)/n}} \\
%=&
=
\frac{2^{3/2} \epsilon^2 G_\alpha}{n C_e C_\alpha \R^{3+(b-2)/n} \left(\delta^{\widetilde d/\widetilde \alpha} k_x\right)^{\widetilde \alpha}}
\int_0^\iy \frac{\mathcal{F}\left( \omega \right)\mathrm{d}\omega}{\omega^{1+(2-b)/n}}.
\end{align}
For the spectrum $\widehat E_\alpha$ to depend only on $k_x \delta $, we set $\widetilde \alpha = \widetilde d$.
It follows that $c=1$. 
%\fcolorbox{black}{yellow}{$c=1$}
%\fcolorbox{black}{yellow}{$\widetilde \alpha = 2+(b-2)/n$}. 
The values $c=1$ and $d=1$ respect the inequalities \eqref{eq:inequalities} and the relation \eqref{eq:equality} obtained in \S\ref{sec:spectrum} from Townsend's spectrum. The decay constant becomes $\gamma=1$, which also falls within the inequality range predicted by Townsend's theory and is consistent with theoretical and experimental studies \citep{tennekes-lumley-1972,fransson-etal-2005}. 

By using the displacement thickness $\delta_d$ instead of $\delta$, as in the MA01 experiments, the spectrum \eqref{eq:sp-ealpha-4} becomes

\begin{equation}
\widehat E_\alpha (k_x \delta_d)
= \frac{A_\alpha G_\alpha \Omega_\alpha}{{(k_x \delta_d)}^{\widetilde \alpha}},
\end{equation}
where

%---------------
\begin{equation}
A_\alpha = \frac{2^{3/2} \epsilon^2}{n C_e C_\alpha \R^{3+(b-2)/n}} \left(\frac{1.72}{\sqrt{2}}\right)^{\widetilde \alpha}, \quad
\Omega_\alpha = \int_0^\iy \frac{\mathcal{F}\left( \omega \right)}{\omega^{1+(2-b)/n}} \mathrm{d}\omega.
\end{equation}
For $k_x \delta_d<0.04$, the experimental data shown in figure \ref{fig:algebraic-fitting} (left) decay algebraically at a smaller rate than at larger $k_x \delta_d$. For fixed $k_x$ and small $\delta_d$, the spanwise wavelength is larger than $\delta_d$, the spanwise diffusivity is negligible, and the flow is described by the boundary-layer equations. It is then expected that the spectrum behaves differently when the boundary-region equations, used in our theoretical framework, instead describe the flow. 

\begin{table}
  \begin{center}
\begin{tabular}{ |c|c| } 
 \hline
 \(\displaystyle
 \widehat E_\beta (k_z)
= \frac{B_\beta G_\beta \Sigma_\beta}{{k_z}^{\widetilde \beta}}
\)
 &
 \(\displaystyle
 \widehat E_\alpha (k_x \delta_d)
= \frac{A_\alpha G_\alpha \Omega_\alpha}{{(k_x \delta_d)}^{\widetilde \alpha}}
\)
 \\
 \hline
 \(\displaystyle
B_\beta G_\beta \Sigma_\beta = 8.3 \cdot 10^{2}
\)
 &
 \(\displaystyle
A_\alpha G_\alpha \Omega_\alpha = 1.91 \cdot 10^{-5}
\)
 \\
 \hline
 \(\displaystyle
B_\beta = \frac{2^3 \epsilon^2}{\R C_e C_\beta} = 27104
\) & 
\(\displaystyle
A_\alpha = \frac{2^{3/2} \epsilon^2}{n C_e C_\alpha \R^{3+(b-2)/n}} \left(\frac{1.72}{\sqrt{2}}\right)^{\widetilde \alpha}=4.86 \cdot 10^{-7}
\) 
\\
 \hline
 \(\displaystyle
G_\beta \Sigma_\beta = 0.0306
\) & 
\(\displaystyle
G_\alpha \Omega_\alpha = 39.32
\) 
\\ 
\hline
\(\displaystyle
 G_\beta = 
\int_0^\iy \frac{\abs{G(\kt)}^2}{\left(1+\kt^2\right)^{(n+b)/2}} \mathrm{d}\kt =
0.0627
\)
& 
\(\displaystyle
G_\alpha = \int_0^{2 \pi} \abs{G(\cot \theta)}^2 \mathrm{d}\theta = 0.314
\)
\\ 
\hline
\(\displaystyle
\Sigma_\beta = \int_0^\iy \frac{\mathcal{F}(\sigma)}{\sigma^2} \mathrm{d}\sigma = 0.488
\)
&
\(\displaystyle
\Omega_\alpha = \int_0^\iy \omega^{(b-2)/n-1} \mathcal{F}\left( \omega \right)\mathrm{d}\omega = 125.22
\)
\\ 
\hline
\end{tabular}
    \caption{Numerical values of quantities related to the energy spectra $\widehat E_\alpha$ and $\widehat E_\beta$.}
    \label{tab:table}
\end{center}
\end{table}

%==========================================================
\subsection{Parameters of the transverse spectrum $\Phi_t$}

We use the exponents $\widetilde \alpha=2.82$ and $\widetilde \beta=1.55$ in \eqref{eq:e-alpha-fitting}, found from the best-fitting analysis in \S\ref{sec:best-fitting}, to solve the algebraic expressions $\widetilde \beta=b-1+n$, found in \S\ref{sec:e-beta}, and $\widetilde \alpha=2+(b-2)/n$, found in \S\ref{sec:e-alpha}. The four coefficients of the transverse spectrum $\Phi_t$ are

\begin{equation}
  c = d = 1, \quad n = \frac{\widetilde \beta - 1}{\widetilde \alpha - 1} = 0.302, \quad b = \frac{\widetilde \alpha \widetilde \beta + \widetilde \alpha - 2 \widetilde \beta}{\widetilde \alpha - 1} = 2.248.
\end{equation}
Table \ref{tab:table} presents the numerical values related to the energy spectra $\widehat E_\alpha$ and $\widehat E_\beta$.
    
%===============================================
\subsection{The spectral function $\mathcal{F}$}

\begin{figure}
 \begin{center}
  \includegraphics[width=0.65\textwidth]{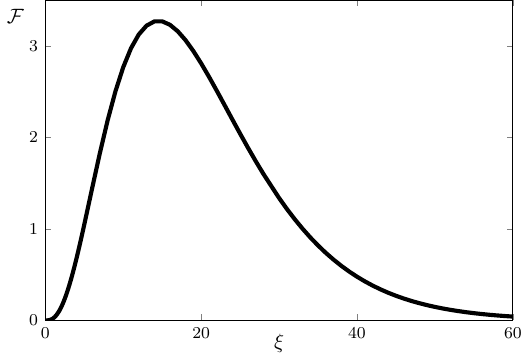}
  \caption{Spectral function $\mathcal{F}(\xi)$ for $A_f=0.03$, $a_1=3$, $a_2=0.3$ and $a_3=0.9$.}
  \label{fig:f-function}
 \end{center}
\end{figure}

A spectral function $\mathcal{F}$ that satisfies the two integrals $\Sigma_\beta$ and $\Omega_\alpha$, given in table \ref{tab:table}, is now chosen. Inspired by \cite{ishihara-etal-2005} and \cite{sagaut-2008}, we select $\mathcal{F}(\xi)= A_f \xi^{a_1}\exp\left(- a_2 \xi^{a_3}\right)$, where the coefficients satisfy

\begin{equation}
\Sigma_\beta 
= \int_0^\iy \frac{\mathcal{F}(\sigma)}{\sigma^2} \mathrm{d}\sigma 
= A_f \int_0^\iy \sigma^{a_1-2} \exp \left(-a_2 \sigma^{a_3} \right) \mathrm{d}\sigma
= \frac{A_f \Gamma\left( \frac{a_1 - 1}{a_3}\right)}{a_3 a_2^{\frac{a_1-1}{a_3}}}
= 0.488,
\end{equation}
%===============
\begin{equation}
\Omega_\alpha 
= \int_0^\iy \omega^{\overline{\omega}} \mathcal{F}\left( \omega \right)\mathrm{d}\omega
= A_f \int_0^\iy \omega^{a_1+\overline{\omega}} \exp \left(-a_2 \omega^{a_3} \right) \mathrm{d}\omega
= \frac{A_f \Gamma\left( \frac{a_1 + \overline{\omega} + 1}{a_3}\right)}{a_3 a_2^{\frac{a_1 + \overline{\omega} + 1}{a_3}}}
 = 125.22,
\end{equation}
with $\Gamma$ the Gamma function, and $\overline{\omega}=(b-2)/n-1$=0.179. We can find multiple combinations of $A_f$, $a_1$, $a_2$ and $a_3$ that satisfy $\Sigma_\beta$ and $\Omega_\alpha$. Figure \ref{fig:f-function} shows an example of the spectral function $\mathcal{F}(\xi)$.

%================================
\section{Conclusions and outlook}
\label{sec:conclusions}

In this paper, we have continued our effort to obtain theoretical and numerical results that explain the experimental findings reported by \cite{matsubara-alfredsson-2001}, one of the most important studies on the impact of free-stream turbulence on the growth and evolution of velocity perturbations in a flat-plate transitional boundary layer. In \cite{ricco-etal-2011}, our theoretical framework and calculations reproduced the main features reported by \cite{matsubara-alfredsson-2001} on the initiation of nonlinear effects within the boundary layer, such as the enhancement of the wall-shear stress with respect to the laminar value, the growth of disturbances in the outer part of the boundary layer and the motion of the peak fluctuations towards the wall. In the present paper, we have instead focused on the collapse of the energy spectral profiles, obtained by \cite{matsubara-alfredsson-2001} when appropriate rescaling was adopted.

The spectral theory of homogeneous temporal-decaying turbulence developed by \cite{townsend-1980} has been utilized to obtain a model spectrum for the streamwise-decaying axial-symmetric free-stream turbulence generated by \cite{matsubara-alfredsson-2001} by use of a grid located in the upstream section of their wind tunnel. Quasi-steady asymptotic solutions of the unsteady boundary-region equations, found by \cite{leib-wundrow-goldstein-1999}, have been used in the analysis of the experimental results of \cite{matsubara-alfredsson-2001}. The quasi-steady approximation was justified by the established finding that the boundary layer acts as a low-frequency-pass filter on the free-stream fluctuations, i.e. low-frequency disturbances are amplified in the boundary layer, while high-frequency disturbances are less prone to reach the core of the boundary layer.

Further work should be directed at measurements of the cross-stream velocity components in the free stream to arrive at a functional form for the transverse spectrum, which is responsible for the generation of the low-frequency Klebanoff modes inside the boundary layer \citep{leib-wundrow-goldstein-1999}. To the best of our knowledge, no experimental data of the free-stream transverse spectrum exist. These data would allow for a better understanding of the response of the boundary layer to the free-stream flow.

As our formulation considers quasi-steady components of the Klebanoff modes, more accurate models that would allow for comparison at any wavenumber and frequency should include perturbations at any value of the scaled wavenumber $\kappa_z$. 
The boundary-layer equations, valid near the leading edge where spanwise diffusion is negligible, should be solved for the cases with $\kappa_z$$\ll$$1$. An evident complication is that the receptivity would then be dictated by the full free-stream spectrum \eqref{eq:spectrum}, which is a combination of the streamwise and transverse spectra, and not only by the leading-order transverse spectrum \eqref{eq:phi-t}. It also follows that velocity components of higher order (with respect to the frequency), such as those appearing in \eqref{eq:M} and the order-one components studied by \cite{wu-dong-2016}, would have to be taken into account.
These improvements could lead to better agreement between the theoretical results and the experimental data at small $k_x \delta_d$ and small $k_z$ in figure \ref{fig:algebraic-fitting}.

In our analysis, only a mild effect of free-stream nonlinearity has been included by modelling the streamwise dependency of the free-stream spectrum, along similar lines to the nonlinear model in \S7.2 of \cite{leib-wundrow-goldstein-1999}. If the streamwise dependency of the free-stream spectrum had not been accounted for, the free-stream decay would have been exponential because dictated by a linearized dynamics, and it would not have been representative of realistic turbulence generated by a grid in a wind tunnel \citep{batchelor-1953}.
Lifting the assumption of low-amplitude disturbances would lead to a better understanding of the boundary-layer response to the free-stream perturbation flow during the nonlinear stages of transition, which may involve secondary instability and the formation of turbulent spots.
An interesting line of research would be the quantitative comparison between such nonlinear receptivity results and experimental data during transition, such as those obtained, for example, by \cite{verdoya-etal-2022}. 

\section*{Acknowledgements}
I would like to express my deepest gratitude to Dr M.E. Goldstein. I give him full acknowledgement for initiating this project, for proposing to use the quasi-steady asymptotic solution of the boundary-region equations and for carrying out the initial analysis on the scaling of $\widehat E_\alpha$ and $\widehat E_\beta$. I thank him for kindly suggesting I should work on this problem. I am also grateful to Dr Elena Marensi for her precious comments and suggestions. I also wish to thank the support of the US Air Force through the AFOSR grant FA8655-21-1-7008 (International Program Office - Dr Douglas Smith) and EPSRC through Grant No. EP/T01167X/1.

\vspace{0.25cm}
\section*{Declaration of Interests}
The authors report no conflict of interest.

%--------
\appendix
\section{Numerical procedures}
\label{sec:num-meth}
The boundary-region equations, given by (5.2)-(5.5) on p. 180 in LWG99 and complemented by the free-stream and initial boundary conditions given by (5.28)-(5.31) on p. 183 and (5.25)-(5.27) on p. 182 in LWG99, are solved numerically. 
\begin{figure} 
  \centering    
  \includegraphics[width=0.6\textwidth]{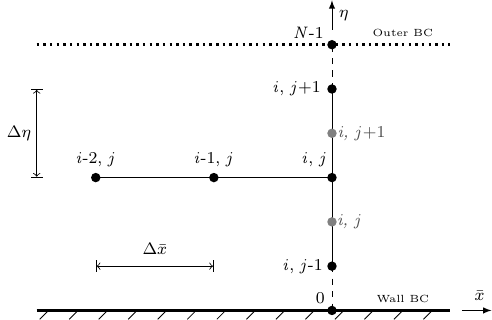} 
  \caption{Sketch of the regular grid (black circles) and staggered grid (grey circles) used for the numerical scheme, adapted from \cite{viaro-ricco-2019}. BC stands for `boundary conditions'.}
  \label{fig:fd-grid}
\end{figure}
As the equations are parabolic along the streamwise direction, a streamwise marching scheme is employed. As shown in figure \ref{fig:fd-grid}, a second-order implicit finite-difference scheme, central in $\eta$ and backward in $\overline{x}$, is adopted, where the derivatives of a velocity component are expressed as 
\begin{equation}
  \frac{\partial q}{\partial \eta} =
  \frac{q_{j+1} - q_{j-1}}{2 \Delta \eta}, 
  \hspace{0.5cm}
  \frac{\partial^2 q}{\partial \eta^2} =
  \frac{q_{j+1} - 2 q_{j} + q_{j-1}}{(\Delta \eta)^2},
  \hspace{0.5cm}
  \frac{\partial q}{\partial \overline{x}} =
  \frac{\frac{3}{2} q_{i,j} - 2 q_{i-1,j} + \frac{1}{2} q_{i-2,j}}{\Delta \overline{x}}.
\end{equation}
If the pressure is computed on the same grid as the velocity components, a pressure decoupling phenomenon occurs. Therefore, the pressure is computed on a grid staggered in $\eta$ as
\begin{equation}
  p = \frac{p_{j+1} + p_{j}}{2}, 
  \hspace{0.5cm}
  \frac{\partial p}{\partial \eta} = 
  \frac{p_{j+1} - p_{j}}{\Delta \eta}.
\end{equation}
The pressure at the wall does not have to be specified and is calculated {\em a posteriori} by solving the $z$-momentum equation at $\eta = 0$.
Due to the linearity of the equations, the system is in the form $\bf{A} \bf{x} = \bf{b}$. For a grid with $N$ points along $\eta$, $\bf{A}$ is a $(N-2) \times (N-2)$ block-tridiagonal matrix where each block is a $4 \times 4$ matrix associated with the four unknowns $\left\{\bar{u}, \bar{v}, \bar{w}, \bar{p}\right\}$. Therefore, the wall-normal index $j$ of the vectors and matrix runs from 1 to $N-2$. The numerical procedure used to solve the linear system is found in~\cite{cebeci-2002} on pp. 260-264. 

%----------------------
\bibliographystyle{jfm}
\bibliography{pr}

\end{document}